\documentclass{optica-article}
\journal{opticajournal} 
\articletype{Research Article}

\usepackage{lineno}
\usepackage{blindtext}
\usepackage{amsmath}
\usepackage{ulem}

\begin{document}

\title{Broadband Diffractive Solar Sail}

\author{Prateek R. Srivastava,\authormark{1} Ryan M. Crum,\authormark{1} and Grover A. Swartzlander, Jr.\authormark{1,*}}

\address{\authormark{1}Chester F. Carlson Center for Imaging Science, Rochester Institute of Technology, Rochester, NY 14623, USA}

\email{\authormark{*}grover.swartzlander@gmail.com} 


\begin{abstract*} 
\textcolor{black}{
The transverse radiation pressure force and acceleration is compared for two parametrically optimized designs: prismatic and two-pillar metasurface gratings.
The numerical results were cross-verified with both Maxwell stress tensor and modal analysis.  Solar blackbody irradiance was assumed for  wavelengths ranging from 0.33 $\mu$m to the grating cutoff at 1.5 $\mu$m, encompassing 83\% of the solar constant.  This multi-objective optimizer study found that neither design comprised of $\mathrm{Si_3N_4}$ performed as well as those corresponding to a low refractive index, low mass density material.  The predicted transverse acceleration of the optimized low-index metasurface grating is compared to that of a state-of-the-art reflective solar sail.
}
\end{abstract*}

\section{Introduction}

The in-space propulsion of sailcraft via solar radiation pressure was originally pioneered by in the 1920s by Tsander and Tsiolkvosy \cite{tsander1924scientific, tsiolkovsky1921extension}. In contrast to rockets which both transport significant amounts of fuel mass and make discrete orbit-changing burns, solar sails can attain extraordinarily high velocities given a low mass and continuous acceleration.
Space organizations such as NASA, JAXA, and the Planetary Society, have improved the technical readiness level of solar sails in recent years, culminating in an assortment of proposed space science missions \cite{johnson2014overview}.
The advent of solar sailing has stimulated advanced concepts that consider the mission objectives as part of the sail design.  For example, missions having a spiral trajectory toward or away from the sun benefit from a sail having an optimal ``lift" force perpendicular to the sun line. To achieve lift a traditional reflective sail must be tilted away from the sun; consequently the maximum lift cannot be achieved owing to the reduced illumination projected area.
In contrast, optical scattering mechanisms like diffraction provide alternative means of transferring photon momentum to the sail in a preferred sun-facing orientation
\cite{Swartzlander2017,Swartzlander2018,Swartzlander2019,Srivastava2019,Srivastava2020,Chu2021_USA,Firuzi2021,Chu2021_China,Swartzlander2022,Swakshar2022,Abdelrahman,Zhang2022,Borgue2023}.
The maximum transverse force on the sail occurs when sunlight is uniformly scattered at 90$^\circ$ with respect to the surface normal of a sun-facing sail.


\begin{figure}[h]
\centering\includegraphics[width=0.5\linewidth]{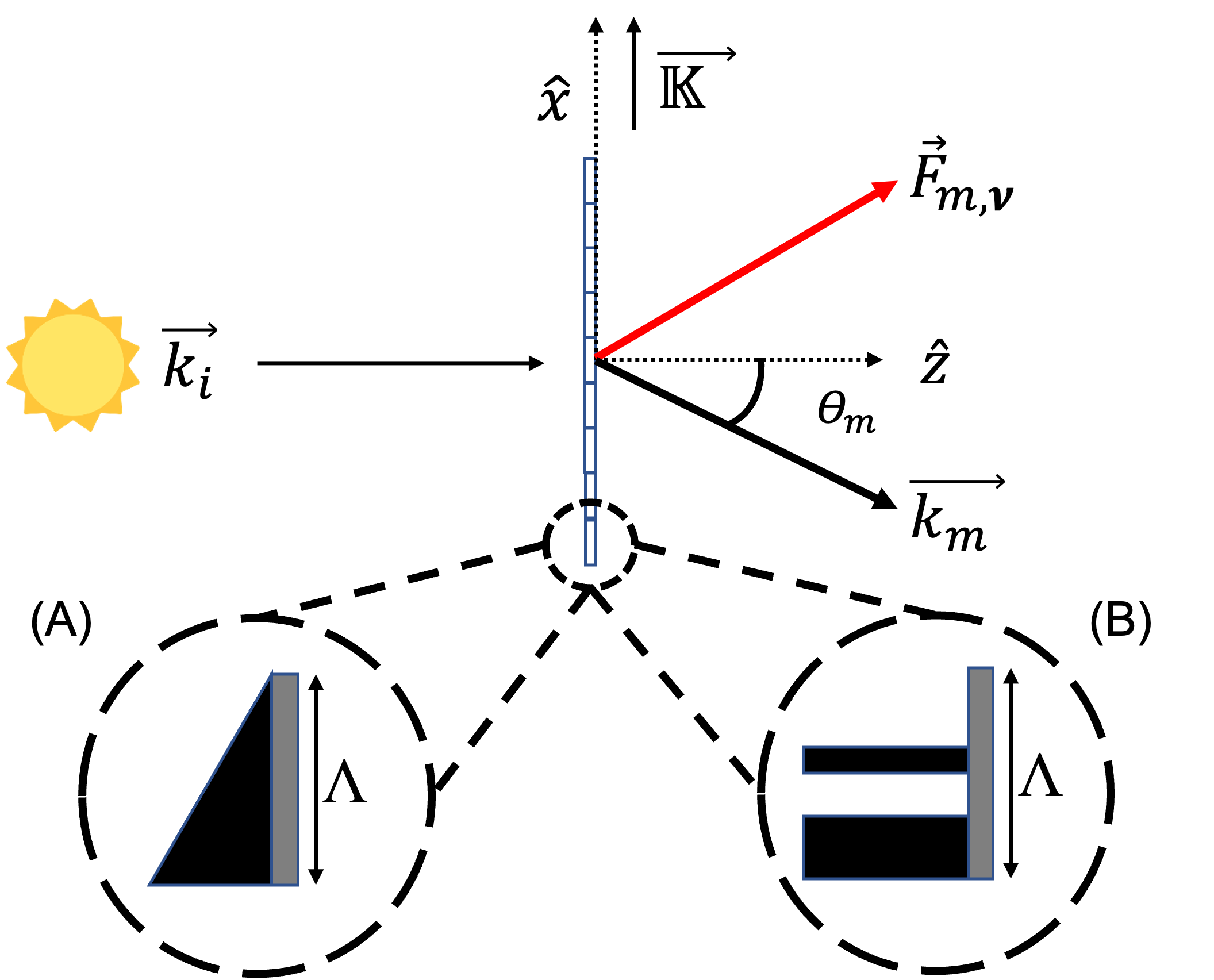}
\caption{Schematic diagram of a solar sail with constituent (A) prism and (B) subwavelength pillar elements of period $\Lambda$ and. The sail diffracts incident light $\Vec{k}_i$ by $\theta_m$ into $\Vec{k}_m$ owing to $\mathbb{\Vec{K}}$, resulting in net radiation pressure force $\Vec{F}$.
}
\label{fig:schematicsolarsail}
\end{figure}

\section{Theory}
\textcolor{black}{To advance the understanding of diffractive sails we explore two designs: a triangular prismatic grating and a metasurface grating comprised of two pillars.
Two material strategies are analyzed for each design.
First we consider an arbitrary non-dispersive dielectric material having a refractive index $n_1$ placed on a thin substrate of index $n_2 = 1.5$.
Finite difference time domain (FDTD) methods are used to account for internal and external reflections of both polarization component of light, and moreover,  the angular scattering distribution across a broad band of optical frequencies.
Likewise, we determine the angular scattering distribution when the grating and thin substrate are made with $\mathrm{Si_3N_4}$.}
The schematic illustration shown in Fig. \ref{fig:schematicsolarsail} depicts a portion of a \textcolor{black}{flat rigid} infinitely periodic grating with period $\Lambda$ in the $x,z$-plane of incidence for a sun-facing configuration, comprised of either (A) prism elements or (B) pillars on a thin substrate.  
\textcolor{black}{Structural flexing and non-normal incidence angle are beyond the scope of this baseline study.}
\textcolor{black}{The grating period $\Lambda = 1.5 \; [\mu \text{m}]$, or equivalently the grating frequency $\tilde\nu = c / \Lambda = 200 [\text{THz}]$
was selected from a consideration of the spectral cut-off condition, the prism mass, and diffraction effects.  
The fraction of blackbody irradiance cut off from diffraction 
decreases with increasing value of $\Lambda$,
whereas the mass of a prism varies as 
$\Lambda^2$.
A large value of the transverse acceleration generally requires negligible spectral cut off and low mass, which combined with a diffraction analysis, provides a value of roughly $\Lambda = 1.5 \; [\mu \text{m}]$.
}

Light is transmitted or reflected light into discreet diffraction angles $\theta_m$
measured with respect to the back surface normal as depicted in Fig. \ref{fig:schematicsolarsail}.
In the reference frame of the sail, the incident and scattered wavelengths are equal, and thus, the respective wave vectors may be expressed 
$\vec{k}_{i}= k\hat{z}$ and 
$\vec{k}_{m}= k (\cos \theta_m \; \hat{z} + \sin\theta_m \; \hat{x})$, where $k = 2\pi/\lambda$.  The diffraction angles are governed by the grating equation: $\sin\theta_m = m\lambda/\Lambda$ assuming normal incidence.
We note that $\cos\theta_m = \pm \sqrt{1-  \sin^2 \theta_m}$,
where $+(-)$ corresponds to transmitted (reflected) light. 
The $m^\text{th}$ order photon momentum transfer efficiency imparted to the sail at the optical frequency $\nu = c / \lambda$ may be expressed 
$\vec{\eta}_{\nu,m} = (\vec{k}_{i} - \vec{k}_{m}) / k$
$= (1-\cos\theta_m)\; \hat{z} - \sin\theta_m \; \hat{x}$,
where $c$ is the speed of light, and normal incidence is assumed.
For a light source having a spectral irradiance distribution $\tilde{I}(\nu)$ the net momentum transfer efficiency $\vec{\eta}$ may be found by integrating over all frequencies and summing over all allowed diffraction orders for both polarization modes
\cite{Swartzlander2022}.
For an unpolarized source like the sun, we assume the spectral irradiance is equally divided into $s$ and $p$ polarization states.
\begin{figure}[t]
    \centering
    \includegraphics[width=0.8\linewidth]{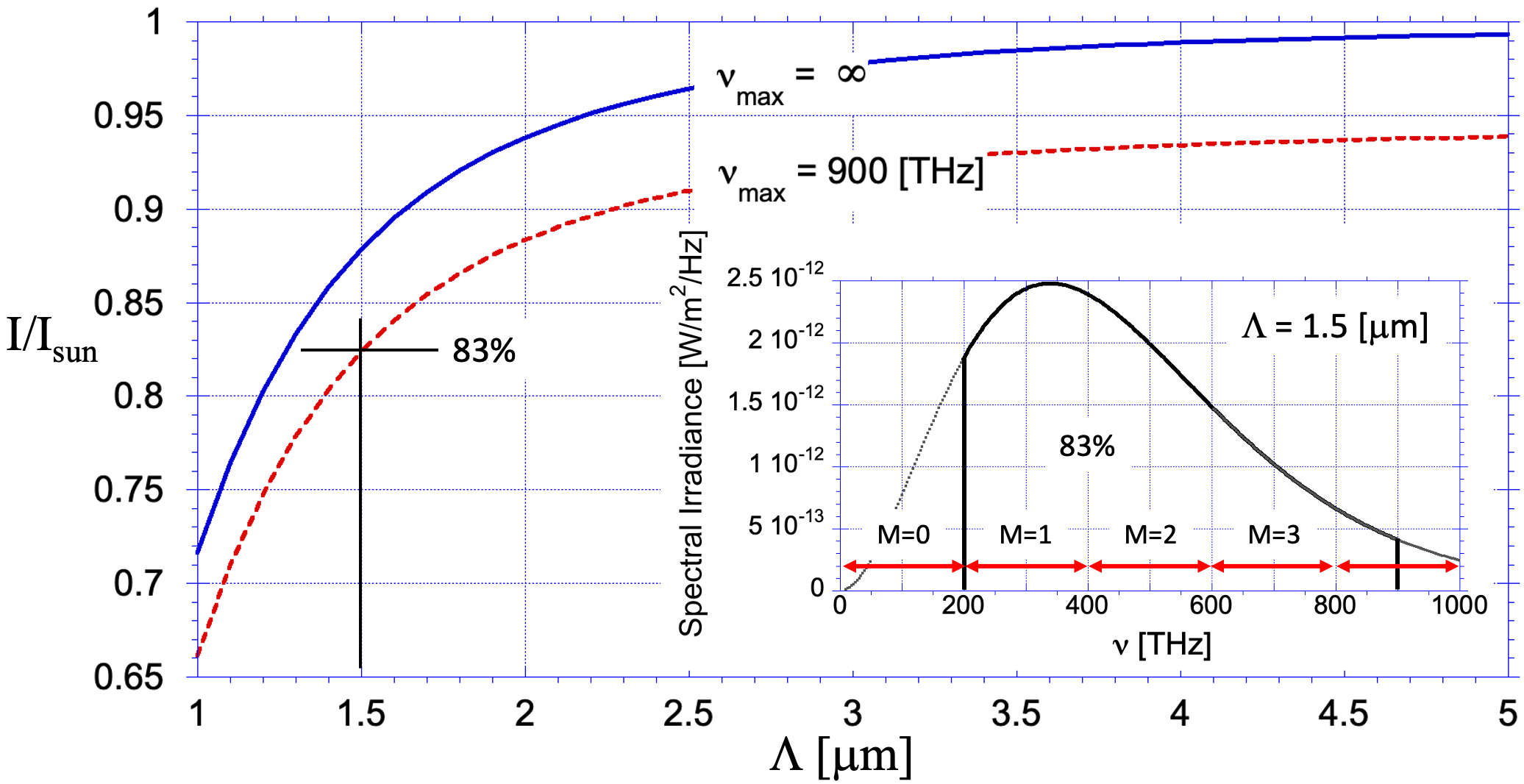}
    \caption{Fraction of integrated solar black body spectral irradiance for the range $\nu_{min} = c/\Lambda$ to $\nu_{max}$, where $I_{sun} = 1360 \; \mathrm{[W/m^2]}$. 
    Insert: Black body spectral irradiance with 
    range $\nu_{min} = 200 \; [\text{THz}]$ $(\Lambda = 1.5 \; [\mu \text{m}])$ and $\nu_{max} = 900 \; [\text{THz}]$ ($0.83 I_{sun}$). Arrows: Range of maximum mode number $M$.
    }
    \label{fig:I0}
\end{figure}

The net radiation pressure force on the sail may be expressed 
$\vec{F} = F_0 \vec{\eta}$, where $F_0 = I_0 A / c$
where $A$ is the sail area and $I_0$ is the irradiance.
For example the solar blackbody irradiance between $\nu_\text{min}$ and $\nu_\text{max} $
of a band-limited blackbody source a distance $r$ from the sun may be expressed

\begin{equation}
I_0 = \frac{R_S^2}{r^2}
    \int_{\nu_\text{min}}^{\nu_\text{max}} 
     \tilde{I}(\nu) d\nu
     = \frac{R_S^2}{r^2} \frac{2 \pi h}{c^2}
    \int_{\nu_\text{min}}^{\nu_\text{max}} 
     \frac{\nu^3 \; d\nu } {\exp(h\nu/k_B T)-1}
    \label{eq:irradiance_band}
\end{equation}

\noindent
where $R_S = 6.957\times10^8$ [m] is the solar radius,  $h = 6.626 \times 10^{-34} \mathrm{[J\cdot s]}$ is the Planck constant, $k_B=1.381 \times 10^{-23} [\mathrm{J/K}]$ is the Boltzmann constant, and we assign $T = 5770.2$ as the effective absolute temperature of the sun. 
Below we assume $r$ corresponds to 1 [AU].
The case $\nu_{min,max} = 0, \infty$ corresponds to the so-called solar-constant, $I_{sun} = 1360 \; [\text{W/m}^2]$.
Values of $I_0$ are plotted in Fig. \ref{fig:I0} as a function of the grating period for 
$\nu_{min} = \tilde\nu = c/\Lambda$ and two different values of $\nu_{max}$: $\infty$ (blue line)
and $900 \; [\text{THz}]$ (red line). 
The case used for our FDTD model,
$\lambda_{min} = 0.333 \; [\mu \text{m}]$ and
$\lambda_{max} = \Lambda = 1.5 \; [\mu\text{m}]$
($\nu_{min} = 200 \; [\text{THz}]$, and 
$\nu_{max} = 900 \; [\text{THz}]$)
includes up to four diffraction orders and spans
83\% of the solar spectrum.
Although wider bandwidths are of interest, FDTD
run times become prohibitively long.

Following Ref \cite{Swartzlander2022} the net radiation pressure force on the sail owing to a band-limited source may be expressed
\begin{equation} \begin{split}
\Vec{F}^{s,p} = \frac{A}{c}
     \int_{\nu_\text{min}}^{\nu_\text{max}}  
    \sum_{m=M_{\nu}^{-}}^{M_{\nu}^{+}}
          \tilde{I}^{s,p}_{m} (\nu) 
          \left( 
          (1 - \cos\theta_m ) \; \hat{z} -\sin\theta_m \; \hat{x}
        \right)
\mathrm{d}\nu
\label{eq:Fmodes2}
\end{split} \end{equation}
\begin{figure}
\centering\includegraphics[width=0.6\linewidth]{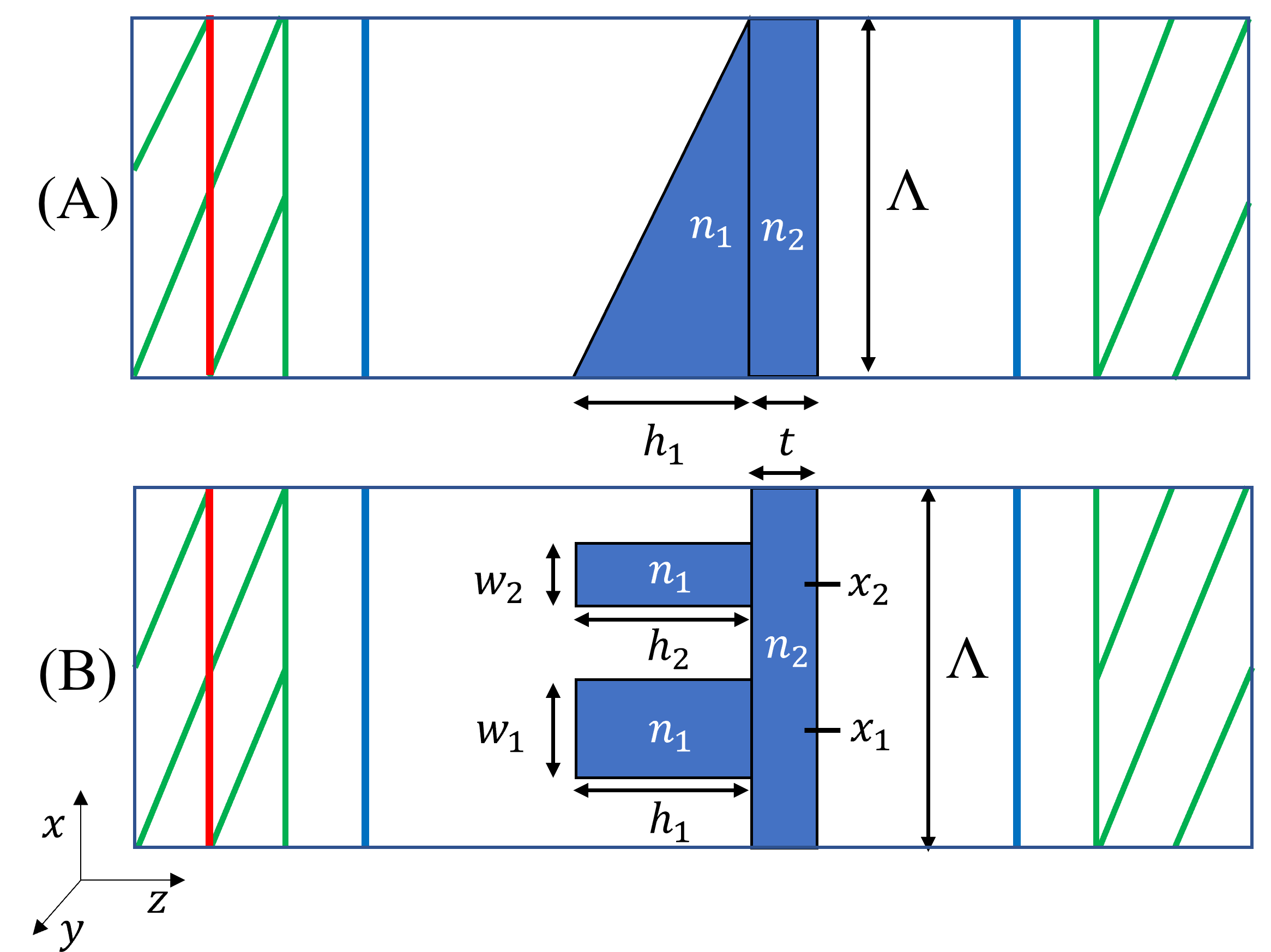}
\caption{
FDTD Schematic: Unit cell of period $\Lambda$ of (A) prism and (B) meta gratings with plane wave source (red line), field monitors (blue lines), and perfectly absorbing boundary layers (green areas).
}
\label{fig:schematicofperiodicarray}
\end{figure}

\noindent where $\tilde{I}^{s}_{m} (\nu)$ and $\tilde{I}^{p}_{m} (\nu)$ respectively correspond to the value of the spectral
irradiance scattered into the $m^\text{th}$ diffraction
order for the $s$ and $p$ polarization states, and
where $\theta_m$ depends on frequency owing to the grating equation which may be expressed, $\sin\theta_m = m c/\nu \Lambda$.
The frequency-dependent cut-off mode numbers at the normal incident are given by 
\textcolor{black}{$M_{\nu}^{\pm} = \pm \mathrm{INT}[\nu / \tilde{\nu}]$ 
(or equivalently $\pm \mathrm{INT}[\Lambda / \lambda]$) }
where $\mathrm{INT}$ represents the integer value of the argument rounded toward zero.
In a lossless system having no guided surface waves that extend to infinity, we expect
\begin{equation}
\tilde{I} (\nu) = 
\sum_{m=M_{\nu}^{-}}^{M_{\nu}^{+}}
\left( \tilde{I}^{s}_{m} (\nu) + \tilde{I}^{p}_{m} (\nu) \right)
\end{equation}
In general $\tilde{I}^{s}_{m} (\nu) \neq \tilde{I}^{p}_{m} (\nu)$ owing to polarization-dependent scattering.

The Maxwell stress tensor $\overline{\overline{T}}_\nu$
may be evaluated at each frequency
as an alternative method to evaluate the net force $\Vec{F}$:

\begin{equation}
    \Vec{F}^{s,p} = 
    \int_{\nu_\text{min}}^{\nu_\text{max}}    
    \Vec{F}_\nu^{s,p} d\nu = 
    \int_{\nu_\text{min}}^{\nu_\text{max}}
    \left( \oint_{S} \overline{\overline{T}}^{s,p}_{\nu, ij}\cdot  d\Vec{S} \ \right) d\nu
    \label{eq:FduetoMST}
\end{equation}
where $S$ is an arbitrary surface enclosing the sail, $\text{d}\Vec{S}$ is the elemental area vector 
and
\begin{equation}
    \overline{\overline{T}}_{\nu,ij}^{s,p} = \epsilon_0 (E^{s,p}_{\nu,i}E^{s,p}_{\nu,j}-\frac{1}{2}|E^{s,p}_{\nu}|^2\delta_{ij}) + \frac{1}{\mu_0} (B^{s,p}_{\nu,i}B^{s,p}_{\nu,j}-\frac{1}{2}|B^{s,p}_{\nu}|^2\delta_{ij})
    \label{eq:stresstensor}
\end{equation}
where $\epsilon_0$ and $\mu_0$ are respectively the vacuum permittivity and permeability, $E$ and $B$ are respectively electric and magnetic field amplitudes, and $\delta_{ij}$ is the Kronecker delta function. 
For a structure that is periodic in the plane of incidence as depicted in Fig. \ref{fig:schematicofperiodicarray} and extended over a distance $L_y$ out of the plane, the only elemental areas that contribute to \eqref{eq:FduetoMST} are 
$d\vec{S}_{z=\pm z_0} = \pm dx \; dy \; \hat{x} \pm  dx \; dy \; \hat{z}$.
The force exerted across the area $L_y \times \Lambda$ of
an infinitely period grating may therefore be expressed
\begin{equation} 
\begin{split}
    \Vec{F}^{s,p} & =  
    \int_{\nu_\text{min}}^{\nu_\text{max}}
    \left( \int_{\Lambda L_y} 
    \left(
    (\overline{\overline{T}}^{s,p}_{\nu, ij}\cdot d \vec{S})_{z=-z_0}
    +
    (\overline{\overline{T}}^{s,p}_{\nu, ij}\cdot d\vec{S})_{z=+z_0}
    \right) 
    \right) d\nu \\
    & = L_y \int_0^\Lambda \big(
    (-T_{xx} - T_{zz} )_{z=-z_0} + 
    ( T_{xx} + T_{zz} )_{z=+z_0} \big) dx
    \label{eq:FduetoMST2}
\end{split}
\end{equation}
\noindent
where $z_0$ is an arbitrary distance from the grating,
and the final integral includes the frequency-integrated stress tensor components $T_{xx}$ and $T_{zz}$.

\section{Numerical Methods}
We used the open source FDTD numerical solver MEEP \cite{Oskooi2010} to solve Eq.s \eqref{eq:FduetoMST} - \eqref{eq:FduetoMST2}, making use of fast built-in ``methods" like \texttt{ForceSpectra} to calculate forces in a specified \texttt{ForceRegion}.  To cross-validate the force values we randomly compared them to values obtained using Eq.\eqref{eq:Fmodes2}, this time using diffraction mode options in MEEP. 
In both cases, Bloch periodic boundary conditions were employed.
The power spectrum of a broadband source in MEEP is defined as the distribution function \texttt{GaussianSource(fcen, fwidth)} where \texttt{fcen} 
and \texttt{fwidth} are respectively the center and width
of the Gaussian distribution. 
Force calculations are made in the frequency domain and we scaled them to correspond to the solar blackbody spectral irradiance.
The red line in Fig. \ref{fig:schematicofperiodicarray} depicts a planar light source propagating in the $\hat{z}$ direction. The blue lines represent so-called monitors where the electromagnetic fields $\vec{E}_\nu^{s,p}$ and $\vec{B}_\nu^{s,p}$ are evaluated for the determination of the Maxwell stress tensor, and where alternatively the spectral irradiance $\tilde{I}_{m}^{s,p} (\nu)$ may be determined to evaluate Eq. \eqref{eq:Fmodes2}. 
The green lines in Fig. \ref{fig:schematicofperiodicarray}
represent perfectly matched layers.  
The square numerical grid elements were set to
$\delta x$ = $\delta z$ = 20 [nm].
The simulation ran until either $E_z$ or $H_z$ decayed to $10^{-6}$ of the peak value. 
\begin{table}[b]
\centering
\caption{Multi-Objective Optimization Scheme:
Nine variables, three objectives, and four constraints.
} 
\renewcommand{\arraystretch}{1.1}
\begin{tabular}{cc}
\hline
$\mathbf{x \in}$ & $ [x_{1,2}, w_{1,2}, h_{1,2}, n_{1,2}, t]$\\
$\mathbf{max}: $ & $F_x^s(\mathbf{x}), \; F_x^p(\mathbf{x})$ \\
$\mathbf{min}: $ & $\mathrm{mass} (\mathbf{x})$ \\
$\mathbf{such \ that:}$ & $1.5 \leq n_{1,2} \leq 3.5$\\
$\mathbf{such \ that:}$ & $-\Lambda/2 \leq x_{1,2} \leq \Lambda/2$\\
$\mathbf{such \ that:}$ & $0 \leq w_{1,2}, h_{1,2} \leq \Lambda$\\
$\mathbf{such \ that:}$ & $0.1\mu m \leq t \leq 0.5 \mu m $ \\
  \hline
\end{tabular}
\label{table1} 
\end{table}

The focus of this study was to determine optimized parameters of the two structures depicted in Fig. \ref{fig:schematicofperiodicarray}, both having the same period $\Lambda = 1.5 \; [\mu\text{m}]$:  
(A) a prismatic grating and substrate having four optimization parameters $n_1$, $n_2$, $h_1$, and $t$;
and (B) a metasurface comprised of two pillars and a substrate having nine optimization parameters
$n_1$, $n_2$, $h_1$, $h_2$, $w_1$, $w_2$, $x_1$, $x_2$, and $t$.
We employed a multi-objective optimizer NSGA-II (with 40 agents, 40 offspring, 150 generations) 
\cite{deb2002fast} with the range of parameter values listed in Table \ref{table1}.
The objectives are to achieve the largest values of transverse force for both polarizations and to minimize the mass.
A representative set of 40 solutions (called Pareto-optimal) were obtained. 
The same procedure was followed for silicon nitride $(n_{\mathrm{Si_3N_4}})$
structures, but in this case $n_1 = n_2$ and $h_1 = h_2$.
Silicon nitride is relatively stable in a space environment, 
its optical properties are well characterized, and its 
lithographic fabrication techniques are mature.
The refractive index $n_{\mathrm{Si_3N_4}}$ varies from 
$ \sim 2.00$ at $200$ [THz]
to $\sim 2.15$ at $900$ [THz]
\cite{Luke_2015}:
\begin{equation}
    n_{\mathrm{Si_3N_4}}^2-1=\frac{3.0249\lambda^2}{\lambda^2-0.1353406^2}+\frac{40314\lambda^2}{\lambda^2-1239.842^2}
\end{equation}

A solar sail is typically used to achieve a spiral trajectory toward or away from the sun.  In this case, the flight time may be minimized when the transverse (lift) component of acceleration
$F_x / M_\text{sc}$ is a maximum, 
where $M_{sc} = m_\text{sail} + m_\text{pl}$ is the total mass of the sailcraft,  $m_\text{sail}$ is the mass of the diffractive sail material, $m_\text{pl}$ is the mass of the payload and structural support mechanisms, and $F_x = F_x^s + F_x^p$.  The transverse acceleration is optimized when both $F_x^s$ and $F_x^p$ are maximized and $m_\text{sail}$ is minimized.  
The sail mass of our two designs may be expressed
\begin{subequations} 
\begin{equation}
m_\text{sail}^\text{prism}  = 
\left( \frac{1}{2} \rho_1  h + \rho_2 t \right)
 N_x^2 \Lambda^2 = \left( \frac{1}{2} \rho_1  h + \rho_2 t \right)  A \\
\end{equation}
\begin{equation} \begin{split}
m_\text{sail}^\text{meta} & = 
 \rho_1 (N_x w_1 h_1 + N_x w_2 h_2) N_x \Lambda +
 \rho_2 N_x^2 \Lambda^2 t \\
& =  \left( \rho_1 w_1 h_1 / \Lambda +  \rho_1 w_2 h_2 / \Lambda  +
\rho_2  t \right) A
\end{split} \end{equation}
\end{subequations}
where $N_x$ is the number of grating periods across the sail,
and $A$ is the area of a square sail.
Ignoring the payload mass ($m_{pl} = 0$) and writing the transverse component of force $F_x = I_0 A \eta_x / c = m_\text{sail} a_x$
we obtain the transverse acceleration for our unladen structures:
\begin{subequations}
\begin{equation}
\label{eq:aprism}
a_x^\text{prism} = \frac{I_0}{\alpha c} \;
    \frac{\eta_x}{\frac{1}{2} n_1  h + n_2 t } 
\end{equation}
\begin{equation}
\label{eq:ameta}
a_x^\text{meta} = \frac{I_0}{\alpha c} \;
    \frac{\eta_x}
    {n_1 (w_1 \mathsf{f}_1 + w_2 \mathsf{f}_2) + n_2  t}
\end{equation} 
\label{eq:accel}
\end{subequations}
\noindent
where $\mathsf{f}_{1,2} = h_{1,2}/\Lambda$ is the fill factor,
and for convenience we associate the refractive index and mass density with a proportionality factor $\alpha$:
$\rho_{1,2} \equiv \alpha n_{1,2}$.
Using the space qualified polyimide material CP1 \cite{CP1} as an example, with 
a specific gravity $\text{s.g.} = 1.54$ and a mean refractive index of 1.57  we obtain $\alpha = 0.98 \times 10^3 \; [\text{kg/m}^3]$. 
For our silicon nitride structures we instead combine
its specific gravity, $\text{s.g.} = 3.17$
\cite{haynes2016crc} with the mean index, $2.02$, to obtain 
$\alpha = 1.57 \times 10^3 \; [\text{kg/m}^3]$.
As seen in Eq. \ref{eq:accel} the transverse acceleration
is independent of the sail area and is implicitly dependent on the grating period $\Lambda$ via the efficiency factor $\eta_x$ (which is found by numerically determining the transverse force $F_x$ value).


\begin{table}[b]
\caption{Optimized parameters and cost function values for (A) prism and (B) meta gratings of arbitrary dispersionless materials, and (C) prism and (D) meta gratings for $\mathrm{Si_3N_4}$, each with period $\Lambda = 1.5 [\mu\text{m}]$, $L_y=1$ [m], $L_x=N\Lambda =1$ [m], $A=L_xL_y$. } 
\centering
\renewcommand{\arraystretch}{1.1}
\small
\begin{tabular}{ccccc}
\hline
Parameters & A & B & C & D \\
  \hline
$h_1 [\mu$m] & 0.76  & 1.12  & 1.02 & 0.62\\
$h_2 [\mu$m] & - & 1.26  & - & $h_1$\\
$w_1 [\mu$m] & - & 0.32  & - & 0.16\\
$w_2 [\mu$m] & - & 0.16  & - & 0.24\\
$x_1 [\mu$m] & - & 0.06 & - & 0.38\\
$x_2 [\mu$m] & - & 0.44 & - & 0.1\\
Prism Angle & 26.9$^\circ$ & - & 34.2$^\circ$ & - \\
$n_1$ & 2.43 & 1.55 & $\text{Si}_3\text{N}_4$ & $\text{Si}_3\text{N}_4$\\
$n_2$ & 1.5 & 1.5 & $\text{Si}_3\text{N}_4$ & $\text{Si}_3 \text{N}_4$\\
$t [\mu$m] & 0.1  & 0.1  & 0.1 & 0.11\\
Force [nN]  & 785  & 787  & 722 & 416\\
mass [$\times 10^{-3}$ kg]  & 1.07 & 0.73 & 1.93 & 0.84\\
 $a_x$ [$\mu$m/$\mathrm{s}^2$] & 731  & 1080  & 373 & 494\\
\hline
\end{tabular}
\label{table2} 
\end{table}

\section{Results \& Analysis}
\begin{figure}
\centering\includegraphics[width=0.7\linewidth]{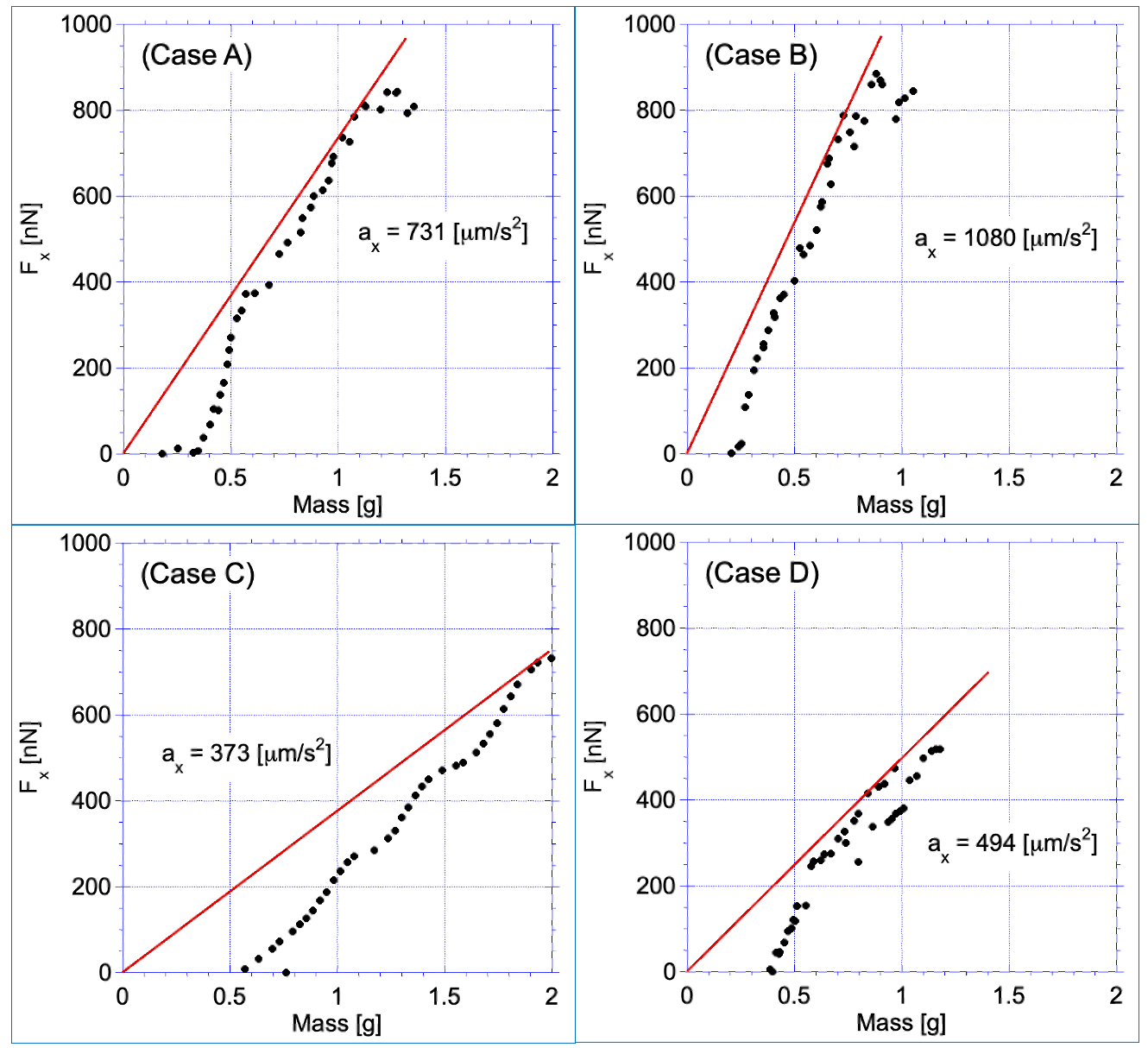}
\caption{Pareto optimal solutions for (A) prismatic and (B) metasurface gratings having arbitrary refractive indexes, and for (C) prismatic and (D) metasurface gratings comprise of silicon nitride. A sun-facing square sail of area $1 [\text{m}^2]$ illuminated with a band-limited solar black body is assumed. The optimal transverse acceleration $a_x$ for each case is determined from the slope of the straight line, and the corresponding design parameter values for the intersecting points are given in Table \ref{table1}. }
\label{fig:paretoptimal}
\end{figure}

Forty representative Pareto-optimal solutions are plotted in
Fig.\ref{fig:paretoptimal} for the two gratings having nine 
arbitrary parameters (A) and (B), and for the two gratings comprised of $\mathrm{Si_3N_4}$ (C) and (D).
The net transverse radiation pressure force $F_x$ is plotted against the total mass of the sail, $m_{\text{sail}}$. In all cases a trend in the data appears: Higher mass sails provide higher forces. To select the most optimal design for each structure we use the greatest value of the transverse acceleration $a_x = F_x / m_\text{sail}$ as the deciding factor (see straight line in Fig.\ref{fig:paretoptimal}).
The parameters for the Pareto-optimal solution that intersects this line are
tabulated in Table \ref{table2} for the four different cases. 

We find that both the prismatic and metasurface structures having arbitrary refractive indexes are able to produce large values of $F_x$, as is evident in Fig.\ref{fig:paretoptimal} for Case A and Case B.  However, owing to the lower mass
of the metasurface structure, its optimal acceleration $a_x = 1080 \; [\mu\text{m}/s^2]$
is $48 \%$ greater than that of the prism grating.  The ${\mathrm{Si_3N_4}}$ structures, Case C and Case D, depict significantly less values of optimized acceleration.
\textcolor{black}{
These values may be compared with a conventional aluminized polyimide sail \cite{McInnes1999} which is roughly $3 \; [\mu\text{m}]$ thick and achieves a momentum transfer efficiency of roughly $90 \%$ of the ideal value of $0.77$ :
$a_x = 680 \; [\mu\text{m}/s^2]$.
This comparison suggests that an optimized metasurface sail is a competitive alternative to a conventional reflective sail.  However, amongst the many unknown fabrication, packaging, unfurling, and space weathering issues is whether a large robust metasurface grating can be fabricated on a thin 
$(< 1 \; [\mu\text{m}])$substrate
\cite{Abdelrahman}.
}

\begin{figure}[t]
\centering\includegraphics[width=0.75\linewidth]{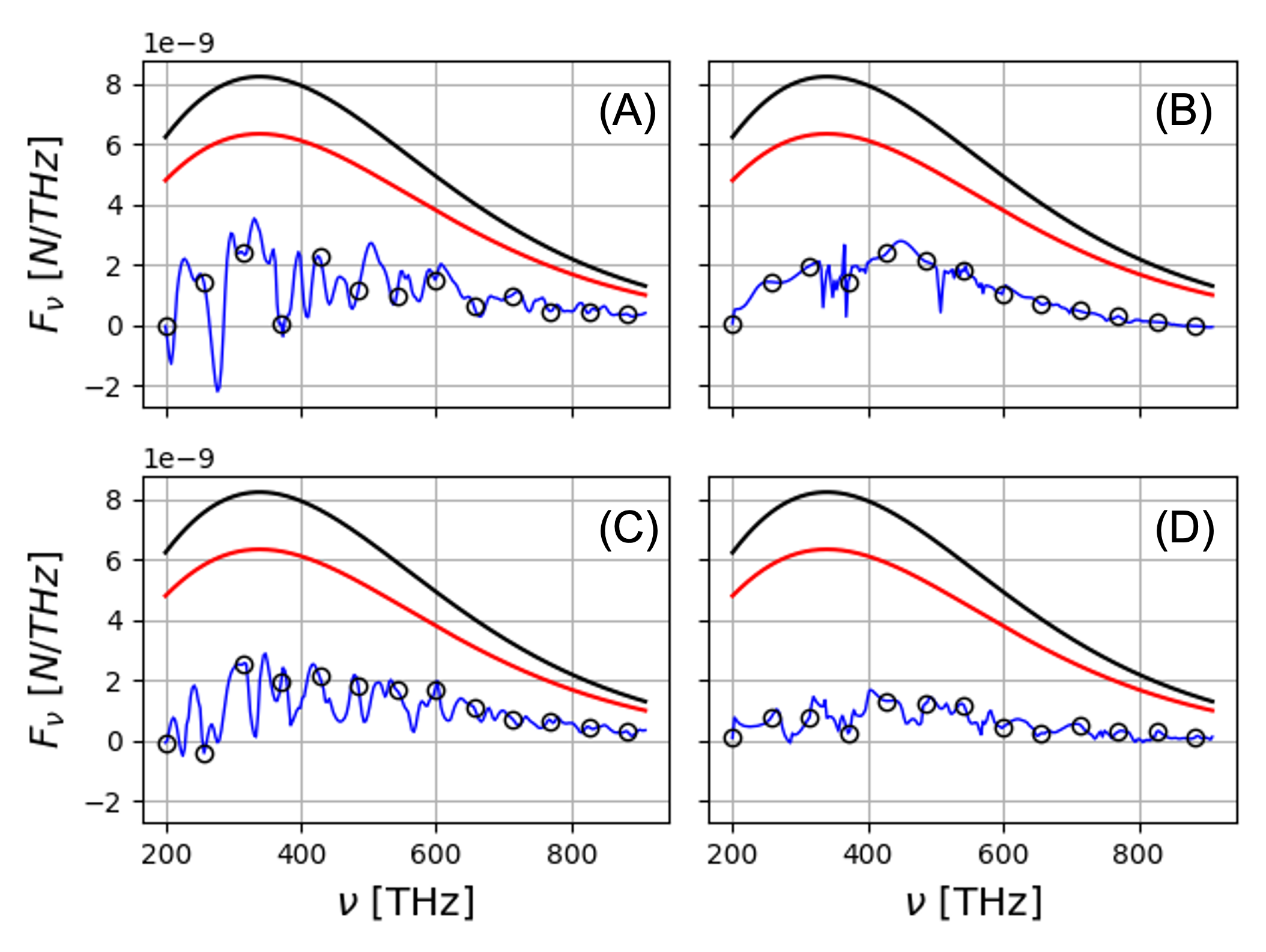}
\caption{Spectral transverse force distribution $F_{x,\nu}$ determined from Maxwell stress tensor (blue line) and modal analysis (open circles) for the four cases described in the text, and for an idealized reflective sail (red line) and the upper theoretical bound (black line). 
An area of $1 \; \mathrm{m}^2$ is assumed.}
\label{fig:MSTvsDE}
\end{figure}

To better understand the spectral force characteristics of the four
sails examined in this study, we plot the transverse spectral force distribution
$F_{\nu,x} = F_{\nu,x}^s + F_{\nu,x}^p$ in Fig. \ref{fig:MSTvsDE}.
The blue line represents the FDTD-obtained values corresponding to the
Maxwell stress tensor calculations, whereas the circles represent the values 
corresponding to our FDTD modal analysis. The excellent agreement between
these two approaches provides a level of cross-validation of the methods.
Fluctuations of the value of $F_{\nu,x}$ are indicative of pronounced 
diffractive variations of the transmitted and reflected light at different optical frequencies, as expected for a small period grating \cite{Swartzlander2017}.  Also plotted in Fig. \ref{fig:MSTvsDE} are theoretical values of 
force for the ideal limit $\eta_x = 1$ (black line) and the ideal reflective sail 
$\eta_x=0.77$ (red line): $F_{x,\nu} = \eta_x \tilde{I}{\nu} A /c$.
\textcolor{black}{
These results suggest that the diffractive sails explored in this study may equal or exceed the acceleration of a reflective sail only if there is a small-mass advantage of the former.  The prism and pillar designs suffer from the effects of external and internal reflections which can scatter light that opposes the desired transverse scattering direction.  For example front surface reflections from the prism in Fig. \ref{fig:schematicofperiodicarray} (A) have positive values of
$k_x$ which oppose the transmitted (refracted) rays.  Those
reflected rays carry 17\% of incident beam power owing to Fresnel reflections.  Less than two thirds of the incident radiation is 
refracted out the back surface owing to internal reflections and shadowing effects from the steep side facets.  It is yet unknown whether the added mass of anti-reflection coatings would provide increased the transverse acceleration.  Other unknowns that are beyond the scope of this paper include the practical limits of assumptions about the rigidity of the sail, the coherence properties of the incident sunlight, and whether the sail can be packaged and unfurled without changing its optical properties.
}

\section{Conclusions}
We performed FDTD simulations coupled with a NSGA-II multi-objective optimizer to determine design parameters for four different grating structures, each having a period of $1.5\; [\mu \text{m}]$ and a sail area
of $1 \; [\text{m}^2]$.
The small grating period was selected to satisfy a small
desired mass and a marginal cutoff wavelength of the solar blackbody spectrum.
Our optimization study included 3 objectives and up to 9 variables, as well as both s and p polarization.
The transverse component of radiation pressure force was determined for a truncated solar black body radiator (200-900 [THz] or equivalently, 0.33 to $ 1.5[\mu \text{m}]$) at 1 [AU] for the purpose of two-orbit changing maneuvers in space.  
An optimized metasurface grating comprised of two pillars per period was found to provide $48\%$ more transverse acceleration than an optimized prism grating owing to the small mass of the former grating.
We found that Silicon Nitride did not perform well for either the prism or two-pillar metasurface design.  Although none of the structures provided radiation pressure force values exceeding those of an ideal flat reflective sail, the diffractive sail may nevertheless provide an acceleration advantage if the proposed sun-facing diffractive sail spacecraft has a total lower mass than a reflective sailcraft. The design of alternatives to flat reflective sails is an emerging area of research
and we therefore believe continued exploration of diffractive designs such as 
hybrid reflective/transmissive structures will provide more efficient solar sails in the future.

\begin{backmatter}

\bmsection{Funding}National Aeronautics and Space Administration(NASA) 
(80NSSC19K0975, 80MSFC22F0165),
Johns Hopkins University Applied Physics Lab (177864).

\bmsection{Acknowledgment}We are grateful to Charles (Les) Johnson and Andrew F. Heaton (NASA George C. Marshall Space Flight Center, Huntsville, AL), and to
Amber L. Dubill (Johns Hopkins Applied Physics Laboratory) for discussions related to solar sailing.
We also thank Rajesh Menon and Apratim Majumder
(U. Utah, Salt Lake City, UT) for meta-material and FDTD modeling discussions.

\bmsection{Disclosures} The authors declare no conflicts of interest.

\bmsection{Data availability} Data underlying the results presented in this paper are not publicly available at this time but may be obtained from the authors upon reasonable request.

\end{backmatter}

\bibliography{sample}
\end{document}